# A Reconfigurable FIR Filter with Memristor-Based Weights

F. Merrikh-Bayat*, F. Alibart, L. Gao, and D.B. Strukov
Electrical and Computer Engineering Department
UC Santa Barbara, Santa Barbara, CA 93105, USA
*farnoodmb@ece.ucsb.edu

*Abstract*— We report on experimental demonstration of a mixed-signal 6-tap finite-impulse response (FIR) filter in which weights are implemented with titanium dioxide memristive devices. In the proposed design weight of a tap is stored with a relatively high precision in a memristive device that can be configured in field. Such approach enables efficient implementation of the most critical operation of an FIR filter, i.e. multiplication of the input signal with the tap weights and summation of the products from taps, in analog domain. As a result, the proposed design, when implemented with fully integrated hybrid CMOS/memristor circuit, is expected to be much more compact and energy efficient as compared to the state-of-the-art approaches.

## I. Introduction

Resistive switching in thin films has many prospective applications in computing [1-7] with the majority of the proposed applications are in digital domain, e.g. in digital memories and reconfigurable digital logic [1, 6, 7]. A very exciting aspect of resistive switching is that very often it is a well-controlled continuous process, allowing for practical implementation of analog memory, which is useful for analog computing. There are already many proposals [2, 3, 4] and even experimental demonstrations of simple circuits taking advantage of analog properties of memristive devices, e.g. to implement tunable gain in operating amplifiers [8], analog memory [9], analog and mixed-signal dot-product operation [10, 11], which can be utilized to implement linear threshold gates [12], conversion circuits [13], and neuromorphic circuits [14, 15]. In this paper, we investigate another very promising analog application of memristive devices – filtering operation. In particular, we investigate several design options for discrete-time finite-impulse response (FIR) filter, present experimental results for a mixed-signal 6-tap FIR filter implemented with breadboard-mounted Pt/TiO$_{2-x}$/Pt memristive devices and off-the-shelf CMOS integrated circuit (IC) components, and discuss performance of the proposed filter implementation. It should be noted that application of memristors in filtering has been proposed before in the context of adaptable RC circuits [16], which is different from what is proposed in this paper. The remainder of this section is devoted to the background information on memristive devices, which are utilized in the experimental demonstration, and FIR filter basics.

### A. Memristive Devices

Figure 1 shows resistive switching curves for a metal-oxide Pt/TiO$_{2-x}$/Pt memristive device, which belongs to the so-called valence-change memory devices according to Waser classification [17]. (The fabrication process for such devices has been described in detail in Ref. [14].) Because of the specific nature of switching, most importantly the presence of negative feedback which slows down the rate of switching for a fixed applied voltage, the reset switching is a fairly continuous and such memristive device can be switched to any

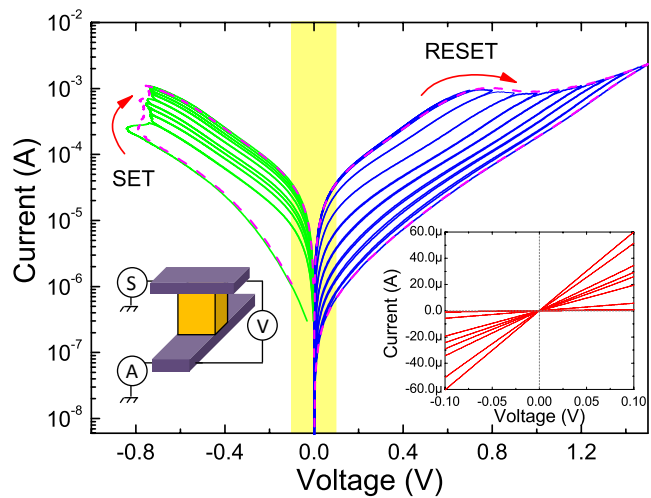

Figure 1. Typical *I-V* switching curves of Pt/TiO$_{2-x}$/Pt memristive device, which are obtained by a quasi-DC triangular voltage sweep followed by a quasi-DC triangular current sweep. The largest reset (set) switching, which is shown with the dashed lines, is obtained with a voltage (current) sweep from 0 to 1.5V (-1 mA), while switching to the intermediate states is implemented with lower stress sweeps with progressively increasing stress amplitude. The inset on the right is zoom-in on a specific voltage range utilized for FIR filter operation (also shown with yellow color on the main figure). The inset on the left shows schematically experimental setup for voltage controlled sweep.

This work is supported by AFOSR under MURI grant FA9550-12-1-0038 and NSF grant CCF-1028336.

intermediate state in the full dynamic range. At the same time, due to highly nonlinear switching kinetics with respect to the applied voltage [18], the state of the device is switched very rapidly (potentially below 1 ns for titanium dioxide devices [19]) with relatively high voltage (e.g. during configuration stage), and retained for a long time (e.g. during operation stage) when smaller voltage biases are applied across the device. Moreover, due to ionic nature of the resistive switching, the aforementioned properties for the memristive devices, i.e. analog switching, high speed and high retention, can be combined with very high device density. Finally, low-temperature fabrication process for the considered devices enables monolithical integration with CMOS circuits [20], which would be necessary for many applications.

For example, applying a simple feedback tuning algorithm [10] a similar Pt/TiO$_{2-x}$/Pt memristive devices with < (20 nm)$^3$ active region have been programmed with ~5-bit and ≲8-bit precision for the crossbar-connected and stand-alone configurations, respectively, even despite large amount of cycle-to-cycle and device-to-device variations upon switching [10, 14]. It is also worth noting that large variations and poor yield, which are the main current challenges for the majority of metal-oxide memristive devices, have been shown to improve upon scaling due to filamentary nature of switching mechanism [7].

*B. Discrete-Time Finite Impulse Response Filter*

The following equation describes the operation of discrete-time *N*-tap FIR filter [21]:

$$y[n] = \sum_{i=0}^{N-1} w_i x[n-i]. \quad (1)$$

Here, $y[n]$ and $x[n]$ are filter output and input, respectively, at sample time *n* and $w_i$ is a corresponding weight of *i*-th tap. Figure 2a shows typical digital FIR filter implementation. Assuming digital input to the filter with $K_d$ bit resolution, the input is fed to *N*-stage-long shift register which preserves the last *N* samples. Each sample is multiplied by a weight tap *w* (stored with $K_w$ bit resolution), which is typically fixed and does not change during filter operation. Finally, the intermediate products of each tap are added up together to create the value of the output signal at sample time $t_n$.

## II. FIR Filter Implementation with Memristor-Based Weights

From Equation 1, it is clear that the filter operation is just a weighted sum or a dot product between input signal and weight vectors. Such dot product operation can be very efficiently implemented with hybrid CMOS/memristor circuits [10, 11] provided that weights do not have to be changed frequently (which is typical for filter operations). This is because, for any reconfigurable circuits, like the one considered in this paper, configuration overhead should be negligible compared to operation time. Because high precision programming of memristive devices is relatively slow process, frequent weight updates may be not practical. Note that the idea of implementing (1) in analog domain has been around for a long time, e.g. with circuits in which weights are realized with floating gate transistors [22, 23].

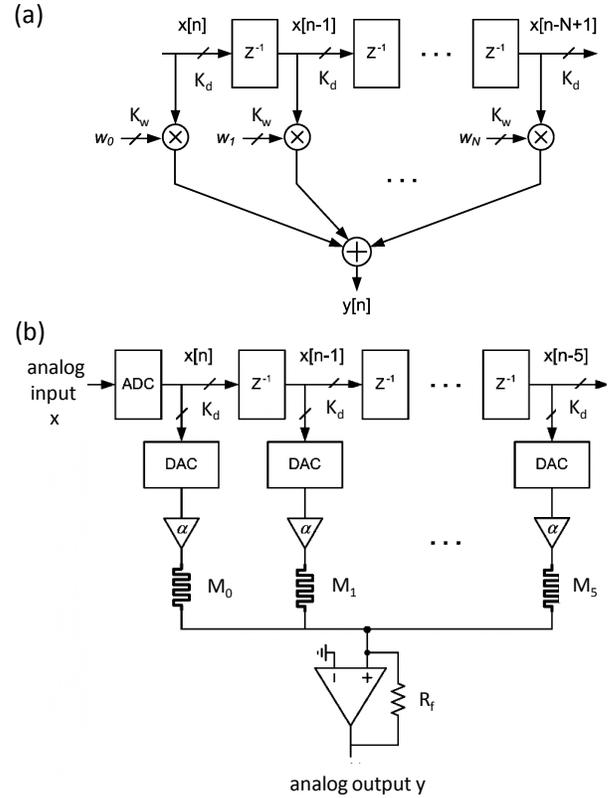

Figure 2. (a) Schematic for digital *N*-tap discrete-time FIR filter and (b) its mixed-signal implementation with hybrid CMOS/memristor circuits with *N* = 6. Here, $K_d$ and $K_w$ are bit resolution for input data samples and filter coefficients, respectively. For experimental demonstration, ADC, DAC, and shift registers are implemented with off-the-shelf IC components ADC0802LCN, DAC0802LCN, and 74LS164N, correspondingly. The value of $R_f$ is 2.2 KΩ. The output of DAC circuits is connected to an opamp-based inverting gain amplifier which is configured to have 0.2V maximum output voltage swing.

Figure 2b shows schematic of experimentally implemented FIR filter circuit in which weights are realized with memristive devices, while summation and multiplication in (1) are performed with conventional operational amplifier (opamp). Because of analog input is used it is first digitized and sampled with frequency *f* with analog-to-digital (ADC). In particular, assuming that physical variables corresponding to *x* and *y* are analog voltages and that opamp is configured as an inverting adder, circuit in Figure 2b implement Equation 1 with

$$w_i = \alpha R_f / M_i, \quad (2)$$

where $M_i$ is the memristance of the corresponding device. The analog gain α is used to force the maximum value of the DAC output to be well below the switching threshold of memristive devices in order to make sure that their configured states are not disturbed during filter operation. Note that static *I-V* characteristics for the considered Pt/TiO$_{2-x}$/Pt devices are roughly linear in the operating range (right inset of Fig. 1) which makes possible analog-input analog-weight dot product operation implementation [11].

## III. EXPERIMENTAL RESULTS

The proposed FIR filter design shown on Figure 2b is experimentally verified with noisy sine-wave input signal and two configurations of weights. In particular, the input signal is formed by adding 0.75V 5Hz sine wave with 20KHz-bandwidth 0.5V-amplitude random noise (Fig. 3a), which is generated by the Agilent 81180A arbitrary waveform generator. In all experiments, $f = 15$KHz and $K_d = 8$ bits.

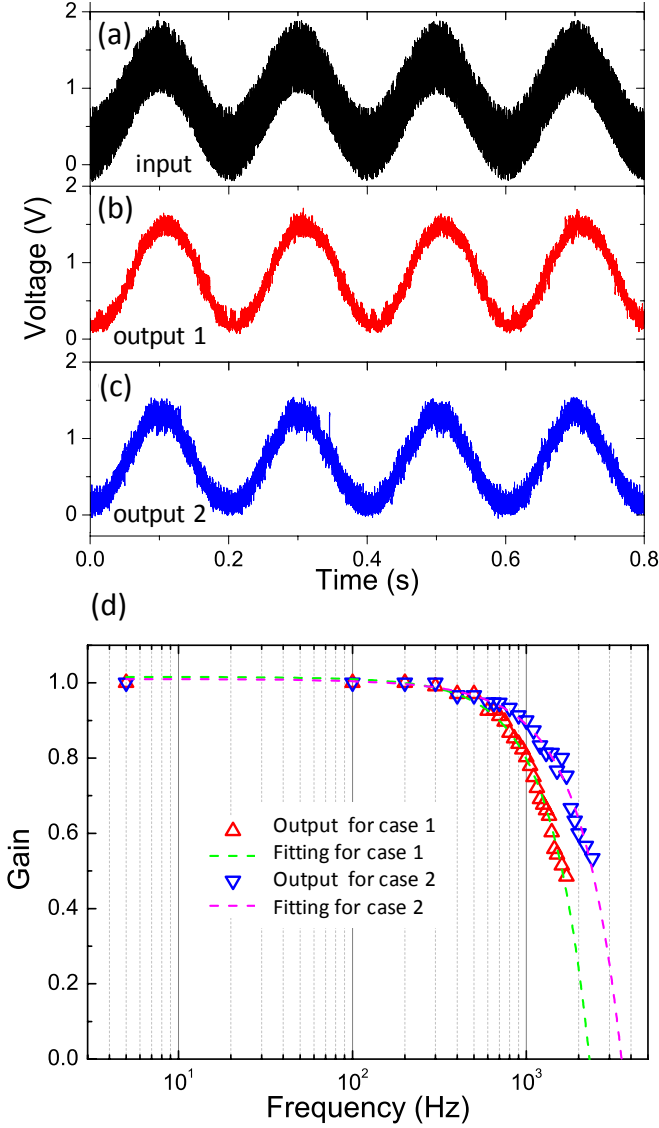

Figure 3. Experimental data for 6-tap hybrid circuit FIR filter (shown on Fig. 2b) which is fed with (a) noisy sine wave input for the two cases: (b) when all weights are set to 2 KΩ ± 5%; and (c) with $M_0 = M_1 = 2$ KΩ ± 5%, $M_2 = 4$ KΩ ± 5%, and $M_3$, $M_4$, $M_5 > 100$ KΩ. Panel (d) shows corresponding frequency response for the two considered cases.

In the first experiment, all 6 memristive devices are programmed to 2 KΩ ± 5%, which crudely corresponds to $K_w = 5$ bits of precision [14]. In this case, FIR filter acts as a simple averaging filter, so that each output sample is an average of the last 6 input samples. Figure 3b shows that amplitude of the noise at the output is dropped by almost factor of 3 as compared to that of input noise, while the amplitude of sine wave remains almost the same, so that signal to noise ratio is improved.

In the next experiment, the weights are programmed to have $M_0 = M_1 = 2$ KΩ ± 5%, $M_3 = 4$KΩ ± 5%, and $M_3 \approx M_4 \approx M_5 > 100$KΩ. Note that since the values of weights of the last three taps are very high, their corresponding contribution to the sum in (2) is almost zero, and only the first three taps are effectively utilized for filtering task. In this case, the amplitude of the noise at the filter output is higher (Fig. 3c) as compared to the previous experiment.

A quantitative comparison of the results for two weight configurations is shown in Figure 3d which shows magnitude of the frequency response. As expected, because half of the weights are effectively zero in one case and the other half is similar for the two considered cases, the cutoff frequency for the second configuration is roughly two times higher as compared to that of the first one.

## IV. DISCUSSION AND SUMMARY

It is worth mentioning that the reason for having conversion from analog input to digital one and then back in demonstrated filter (Fig. 2b) is due to breadboard setup implementation, which limits the number of memristive devices which can be connected on the breadboard and their tuning accuracy and, in general, is not convenient for analog circuits due to large capacitive coupling noise. Such conversion can be implemented more efficiently or eliminated completely with fully integrated CMOS/memristor circuits.

For example, fully analog implementation for the proposed filter design, without ADC and DAC circuits, can be achieved by utilizing analog shift register [24]. Another approach is to combine DAC circuits with dot-product operation of the filter (Fig. 4). In this case, implementation of $N$-tap FIR filter involves $N \times K_d$ memristive devices with $< (K_w + K_d)$ bit resolution. Note that this design relies on digital-input analog-weight computation and suitable for memristive devices with static nonlinear I-V characteristics, as long as the output voltage of digital shift register is lower than the threshold of memristive devices.

Due to breadboard implementation the performance of demonstrated FIR filter is not representative of the potential performance for this approach. Instead, let us try to estimate performance for the case when such filter is implemented with fully integrated CMOS/memristor circuits [20]. In particular, due to very similar circuitry memristor-based design can be compared with that based on floating gate transistors, which is much more energy efficient and compact as compare to conventional approaches for the cases when required precision is not very high [22, 23]. In both approaches memory elements allow to store weights with ~ 1% accuracy [10, 23] when implemented with macroscale (>100 nm) devices. However, the bit resolution for ultimately scaled floating gate transistor is expected to be less due to limited number of electrons which can be stored on a floating gate, e.g. < 4-bit for sub 20-nm features. This is not the case at least for ionic-based memristive devices with filamentary conduction, in which then active area is nanoscale even for macroscale devices. Moreover, memristive devices have smaller footprint and can be indirectly integrated on top of CMOS stack and thus memristor-based design should have better density, latency and energy performance. Note that both design share similar challenges, e.g. high sensitivity to temperature and random telegraph noise [11].

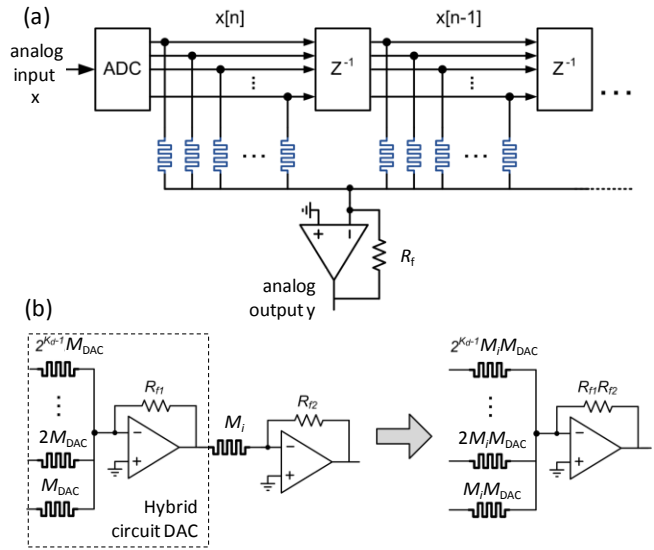

Figure 4. A proposal for (a) more efficient hybrid circuit implementation of mixed-signal FIR filter obtained by (b) combining binary-weighted DAC circuit in each tap with dot-product operation of the filter.

In summary, we have experimentally demonstrated operation of mixed-signal 6-tap FIR filter in which weights are stored with high precision in Pt/TiO$_{2-x}$/Pt memristive devices. Such approach enables efficient implementation of the most critical operation, i.e. multiplication of the tap weights with input signal and summation of resulting products, in analog domain. Due to unique properties of memristive devices FIR filter implemented with fully integrated CMOS/memristor hybrid circuits is expected to have superior performance as compared to state-of-the-art approaches.

ACKNOWLEDGMENT

Authors would like to acknowledge B. Hoskins for help with device fabrication and characterization and K.-T. Cheng and C. Teuscher for useful discussions.